\documentstyle[prb,aps]{revtex}
\begin{document}
\draft
\title{Slow dynamics of stepped surfaces}

\author{Anna Chame\footnote{Permanent address: Instituto de F\'{\i}sica,
Universidade Federal Fluminense, 24210-340, Niter\'oi RJ, Brazil.} }
\address{D\'epartement de Recherche Fondamentale sur la Mati\'ere Condens\'ee,
 Commissariat \`a l'Energie Atomique, 17 Avenue des Martyrs, F-38054 Grenoble
 Cedex 9, France\\ }
\author{Sylvie Rousset}
\address{Groupe de Physique des Solides, Universit\'e Paris 7 et 6, Laboratoire associ\'e au CNRS, 2 Place Jussieu 75251 Paris, France \\}
\author{Hans P. Bonzel}
\address{Forschungszentrum Juelich, Institut f\"ur Grenzflaechenforschung und 
Vakuumphysik, D-52425 Juelich, Germany}   
\author{Jacques Villain}
\address{D\'epartement de Recherche Fondamentale sur la Mati\'ere Condens\'ee,
 Commissariat \`a l'Energie Atomique, 17 Avenue des Martyrs, F-38054 Grenoble
 Cedex 9, France\\
\&  Centre de Recherches sur les Tr\`es basses temp\'eratures, CNRS,
B.P. 166, 38042  Grenoble Cedex 9}

\date{\today}
\maketitle

\begin{abstract}
\tightenlines{ Two kinds of configurations involving steps on 
surfaces are reviewed. The first one results from an initially planar
vicinal surface, i.e.  slightly deviating from a high-symmetry (001) or (111)
orientation. In some cases, these surfaces separate into domains of different
orientations by a mechanism which is very similar to phase separation
in mixtures. The domain size initially increases with time and
goes to a finite limit, whose value is related to elastic phenomena.
 The second kind of configurations results 
from making grooves in a high-symmetry
surface. The surface smoothes out and takes an intermediate shape
with facet-like hills and valleys, which are the source of a
controversy which we try to clarify as much as possible. }

\end{abstract}



\widetext
\section{
Motivation.}
\label{ch1}

Surfaces are of great technological and fundamental interest since 
many processes, such as  crystal growth,  catalytic reactions, production of 
nanostructures,  occur mainly at surfaces.

The surface of an industrial catalyser is too complicated to allow for 
an observation of the microscopic processes. On the other hand, 
a high-symmetry, (001) or (111),  surface of a cubic crystal is too
simple. The only process which occurs is the diffusion of single atoms, which 
cannot easily be observed. 

The present review will be focussed on {\it stepped} surfaces of cubic 
crystals, i.e. surfaces made of large high-symmetry terraces separated by 
steps of thickness one monolayer (Fig. \ref{fig1}) .  

The fundamental interest of stepped surfaces is that they are both simple
and not {\it too} simple. 
The ensemble of the steps can be arranged  
into many different structures, due to various factors which can play 
together or compete among themselves. Simple models can include  
attachment/detachment of atoms to/from the steps,  diffusion or evaporation 
of atoms, and  also step fluctuations. The steps can interact with each other, 
through short  or  long range interactions. The effect of a flux of 
atoms over the stepped surface  can be considered, growth can occur through 
step flow, there can be step bunching, etc. The dynamics of steps gives 
informations 
on the microscopic parameters (energy, diffusion constants...) and are
a test of our ability to solve problems of nonlinear dynamics.

The experimental observation of steps is relatively easy, for instance by 
scanning tunnel microscopy (STM)~\cite{Williams}. The present review will 
not be concerned with isolated steps, but only with experiments in which
interactions between steps has an influence on their motion. Two examples 
are given in the next section.


\section{Experiments}
\label{ch2}

Solids are  generally not in their equilibrum shape, and that is a good thing
since otherwise one could not make, for instance, tools or artwork. 
However, small solids particles 
can reach their equilibrium shape at high enough temperature. 

\subsection{Grooves}
\label{sch2.1}

A case of interest in the present review is that of a crystal surface
in which periodic grooves have been made. Figure 2 shows the example
of periodically modulated Ni(110) and Ni(100) surfaces which have been
annealed at  800  C to obtain a steady state profile shape. This shape is
sinusoidal in the first case but trapezoidal in the second. If
annealed for long times, both profiles will decay and approach a flat
surface. If the profile has a short enough wavelength and the
temperature is high enough (wavelength $ \lambda \approx 10  \mu m$ ,
typical temperatures $T$ between $900-1300 K$) the profile smoothens
after reasonably short times~\cite{Blakely,Yamashita} (Fig.
\ref{fig2}).

The relevant questions in context with this experiment are:
 i) How does the amplitude $2h(t)$ decay?
Exponentially or not? ii) What is the transient shape of the profile?
iii) How does the relaxation time $\tau$ depend on the wavelength
$\lambda$?
Mullins~\cite{Mullins} has given a simple theory
(reviewed in chapter \ref{ch2'}) valid above the roughening transition
(we shall recall later what this means). Mullins' answers to the above
questions, under the assumption of small slopes,  are the following:
i) The decay is exponential.  ii) The profile is sinusoidal or becomes
rapidly sinusoidal if it is initially not. iii) The relaxation time
$\tau$ is proportional to a power of $\lambda$ which depends on the
type of dynamics. Mullins' theory will be reviewed in chapter
\ref{ch2'}.

Early experimental observations~\cite{Yamashita} evidenced that, for
initially prepared sinusoidal profiles on Ni(100), (110) and (111)
single  crystals, the evolution towards  equilibrium was found to be
different depending on which surface the profile is created. For
(110) surfaces the profile maintains the sinusoidal shape during the
decay and the time law is strictly exponential. The relaxation time
scales with the fourth power of the wavelength of the profile. In
contrast, in the case of the high symmetry (111) and (001)  surfaces,
flat regions, reminiscent of facets, form at the maxima and minima.
The decay of these profileswith time is slower than for the ones on
Ni(110)~\cite{Bonzel,Preuss}. Non-sinusoidal profile
shapes were also observed~\cite{Voigtlander}  for the (100) and (111)
surfaces of gold shown in Figure 3. This case, however, is even more
complicated than Ni, because vicinal Au(100) and (111) surfaces are
unstable and facet. This problem will be discussed in parts \ref{sch2}
and \ref{ch11}.


\subsection{Phase separation on vicinal surfaces}
\label{sch2}

	A vicinal or miscut surface is a surface obtained by cutting a crystal 
along a plane making a small angle $\alpha$ with a low index plane, (001) or 
	(111) for instance. The ideal structure of a vicinal surface then consists 
	of high symmetry terraces separated by equally spaced steps
of atomic height. Indeed, such configurations have 
	been observed, for example on vicinal copper surfaces at room 
	temperature~\cite{Girard,Masson}, or on 
	vicinal Si(111) surfaces at elevated temperature~\cite{Alfonso}. 
	Although thermal 
	fluctuations have been studied in detail on vicinal surfaces, and are 
	responsible for a terrace width distribution~\cite{Williams}, 
the average distance 
	between steps is usually consistent with the ideal expected value.

However, it has been  known for a long time~\cite{flytz}  	
that upon thermal treatment, 
	some stepped surfaces are unstable and separate into parts (or `facets')
	of two or three different orientations. This separation is analogous to 
	phase separation of unstable mixtures 
	and the resultant morphology is a hill-and-valley structure. 
Such a situation occurs for instance on Au(111) vicinal 
surfaces~\cite{Pourmir}. Two examples are shown in figures \ref{sylvie1}
and \ref{sylvie2}, namely Au(11,9,9) and Au(5,5,4). 	
Both surfaces are tilted by about 6$^{\circ}$ with respect to the (111) orientation and can be figured out as consisting of [1,-1,0] monoatomic steps 
which are initially equally spaced by about 24\AA. 
Both surfaces are unstable. After a thermal treament in ultra high vacuum, 
a faceted morphology is clearly seen in figures \ref{sylvie1}
and \ref{sylvie2}. In the simplest 
case, the unstable surface evolves towards an array of step bunches 
separated by high symmetry terraces. For instance, a (11,9,9) gold 
surface (Figure \ref{sylvie1}) transforms into (111) terraces of 35 
\AA width, 	separated by small portions of another vicinal surface, namely the 
(7,5,5) orientation. In the case of the (5,5,4) surface 
(Figure \ref{sylvie2}),
the transformation leads to facets which are both portions of vicinals. 
All  steps have still the same sign. The phase with the greatest 
slope corresponds to a (3,3,2) orientation, whereas the other one is 
misoriented from (111) by about 4$^{\circ}$. 
The additionnal long range order 
displayed by the faceted morphologies will be discussed in section XII. 
The stability of the facets may be interpreted as caused by an interaction 
between steps which is not uniformly repulsive, as will be seen in 
subsection \ref{sch3b}. 

Another possible cause of faceting is surface melting~\cite{zozo,HMB}. For instance,  
Frenken et al.~\cite{Frenken} have shown, using medium energy ion scattering, 
that vicinal Pb(111) can separate into two surfaces of non-zero step 
density, a melted one and a dry one.

Measurements~\cite{Phaneuf} of the faceting of vicinal Si(111) surfaces show 
that this surface also rearranges into a hill-and-valley structure. 
At high temperature this surface consisted of an 
uniform density of steps and at low temperatures two ``phases'' appear: 
the ($7 \times 7$) reconstructed (111) facets and the unreconstructed step 
bunches~\cite{Williams}. 
  
 As in ordinary phase separation, the two coexisting
 phases can be deduced by the so-called double tangent 
 construction~\cite{noz,JVAP} if one knows the thermodynamic potential
 as a function of the relevant variable (the concentration in the case of
 a mixture, orientation in the case of a surface). 

 This thermodynamic potential, in the case of a stepped surface, can be 
 in principle deduced, or at least understood, from the knowledge of the
 interactions between steps, which are the subject of section
 \ref{ch3}.


\section{Various types of kinetics and Mullins' theory}
\label{ch2'}

In order to model the experiments reported in Subsection \ref{sch2},
we consider a crystal surface which is approximately 
a plane parallel to 
the $xy$ direction, but for a small initial perturbation which slowly decays in time. We wish to study this decay. We can consider for instance an
initially sinusoidal (unidirectional)  profile of wavelength $\lambda$,
namely:
\begin{equation} 
z(x,t=0) = h(0) \cos {2\pi x \over \lambda} \; .
\label{perfil}
\end{equation}

The important quantity which drives the relaxation process is the chemical 
potential $\mu$, i.e. the thermodynamical potential per particle.
If the surface is out of equilibrium, $\mu$ is not constant on 
the surface and can be related~\cite{Herring,Mullins2}
to the surface free energy density
by the Herring-Mullins formula. This formula can be written 
in a simple way~\cite{JVAP}
if one introduces  the  free energy density $\varphi$ per unit
area of the  projection onto any reference plane, 
e.g. the $xy$ plane:
\begin{equation}
\mu  = \mu_0 - a^{2}
\left[ {\partial \over \partial x} \left({\partial
\varphi \over \partial z_x} \right) +  {\partial \over
\partial y} \left({\partial \varphi \over \partial z_y} \right) \right] \;,
\label{8.3}
\end{equation}
where $z$ is the height of the surface above
the $xy$ plane and $z_\alpha=\partial z/\partial x_\alpha$.
This formula makes sense if  $\varphi$ is an analytical 
function of  $ z_x $ and $ z_y$. If it is so, the surface is called 
{\it non-singular}. 
The remainder of this section is devoted to the simple case of 
a surface which is non-singular in all its orientations. As will be seen
in the next section this is not the case of stepped surfaces, but
the forthcoming study will provide us the opportunity 
to introduce several useful concepts.

For not too high temperatures, atoms can neither evaporate nor diffuse through
the solid, so that  surface relaxation takes place
through diffusion of atoms on the surface. This case is called {\it surface 
diffusion dynamics}. One can then write a conservation equation which expresses the absence of evaporation: 
\begin{equation}
\partial z/ \partial t   = - a^2 \vec \nabla \cdot \vec j_s \;.
\label{diffusion} 
\end{equation}
where $a$ is the atomic distance.

Mullins~\cite{Mullins}
 wrote the surface current density $\vec j_s$ in (\ref{diffusion}) as
\begin{equation}
\vec j_s = - a^{-2} \beta\tilde D_s \vec \nabla  \mu   
\label{current}
\end{equation}
where $\mu$ is the local chemical potential, $\beta$ is $1/(k_BT)$ and 
$\vec \nabla  = 
(\partial \sb{x} , \partial \sb{y}) $. 
The surface ``mass diffusion constant'' $\tilde D_s$ is defined as
$\tilde D_s = a^2 \rho_0 D  $, where  
$\rho_0 $ is the adatom density  at equilibrium and $D$ is 
the adatom diffusion constant (we suppose that the advacancy contribution to diffusion is negligible compared to the adatom contribution).

Mullins~\cite{Mullins} approximated $\varphi$ by a quadratic  
function which will be assumed isotropic for the sake of simplicity. 
This procedure is usual in liquids and yields: 
\begin{equation}
\varphi (z_x,z_y) \simeq  {c^2 \tilde \sigma \over 2} 
\left(z_x^2 + z_y^2 \right) 
\label{8.4}
\end{equation}
where $\tilde \sigma $ is the surface stiffness~\cite{noz,JVAP} and $c$ is 
the interatomic distance along the  $z$ axis.

Relations (\ref{8.3}) and  (\ref{8.4}) yield 
\begin{equation}
\mu  = \mu_0 - a^2 c^2 \tilde \sigma  \left(z_{xx}+z_{yy} \right) \;.
\label{mu}
\end{equation}

Inserting  (\ref{mu}) 
into (\ref{current}) and (\ref{diffusion}), one finds 
\begin{equation}
\partial z/ \partial t  = - a^2 K  \nabla^2 \left(\nabla^2 z \right) \;,
\label{eqmotion}
\end{equation}
where $K = c^2 \beta\tilde D_s \tilde \sigma >0$. This equation of motion is 
linear, so that an initially sinusoidal profile remains sinusoidal:

\begin{equation} 
z(x,t) = h(t) \cos {2\pi x \over \lambda} \;.
\label{8.2b}
\end{equation} 

Using (\ref{eqmotion}), the amplitude $h(t)$ in (\ref{8.2b}) is seen to decay 
exponentially with the time $t$,
\begin{equation}
h(t)=h(0)\exp(-t/\tau)
\label{tau}
\end{equation}

with a relaxation time

\begin{equation} 
\tau (\lambda) = { \lambda^4 \over\ 16 a^2 \pi^4 K}  \;.
\label{8.13}
\end{equation}

For an arbitrary, but periodic initial profile,  the 
shape will anyway become sinusoidal after some time
since it can be regarded 
as a sum of many sinusoids of different wavelengths (Fourier decomposition). 
At long times, only  the component of longest wavelength persists.

The above results hold for surface-diffusion kinetics. 
The comparison with experiment allows it to determine  
systematically  the diffusion constant 
$\tilde D$ on the (110) face of fcc metals~\cite{Blakely,Bonzel}. 

Another case is when matter is carried through the vapour.
This process, important at high temperature, 
is called {\it evaporation-condensation kinetics}.
For simplicity,  the chemical potential of the vapour 
is always uniform and equal to the value $\mu_0$ of the bulk solid. 
The evaporation rate at a point of the surface 
where the chemical potential is $\mu$ is proportional to the difference 
$\mu - \mu_0$ :
\begin{equation}
\dot z \propto  \mu - \mu_0
\end{equation}   
Then using  expression (\ref{mu}) for the chemical potential,  one obtains
\begin{equation}
\dot z = {\cal A}  \left(z_{xx}+z_{yy} \right)
\label{evap}
\end{equation}
where ${\cal A}$ is a temperature-dependent coefficient. Since 
(\ref{evap}) is 
a linear equation, there are again sinusoidal solutions which satisfy 
eqs. (\ref{8.2b})
and (\ref{tau}). However, for  evaporation-condensation  kinetics, 
the above equations yield
$\tau \propto \lambda^2$ in contrast with (\ref{8.13}).

A third way to transport matter,  and thus to allow for the  
relaxation of the profile, is  bulk diffusion. When this process 
dominates, it can  be 
shown~\cite{Mullins} that, in the non-singular case, $\tau \propto \lambda^3$.


\section{Stepped surfaces}
\label{ch2''}

The free energy of a stepped surface is the sum of the free energy of the high 
symmetry parts and the free energy of the steps. The simplest case is that of 
a regular array of straight, parallel steps at distance $\ell$ 
(Fig. \ref{fig6}). If the step energy per unit length is $\gamma$
and if $\ell$ is so large that the interaction between steps can be 
neglected, the  projected free energy density 
resulting from  high symmetry parts is independent of $\ell$  
while the free energy of the steps is $\gamma/\ell$. The total
projected free energy density can be written as

\begin{equation}
\varphi(z_x,z_y)=\varphi(0,0) + \gamma (z_x^2+z_y^2)^{1/2} 
\label{phi0}
\end{equation}
In contrast with (\ref{8.4}), this expression is not analytic in 
$z_x$ and $z_y$ (Fig. \ref{fig6}b). A high symmetry surface is {\it singular}
and   Mullins' theory is not applicable.

More precisely, a high symmetry surface  is singular below its {\it roughening 
transition}~\cite{Beijeren} which occurs at a temperature $T_R$. Above $T_R$,
$\gamma$ vanishes, (\ref{phi0}) does not apply, and one can argue that 
(\ref{8.4}) applies. For the  (001) and (111) faces of most fcc materials, 
$T_R$ is very close or equal to the melting
temperature. Low symmetry surfaces also have roughening transitions, 
which occur at much lower temperatures.

In the case of stepped surface, the three types of kinetics (surface-diffusion,
evaporation-condensation, bulk diffusion) are still possible. However,
surface-diffusion kinetics subdivide 
into two classes \cite{Nozieres,Cahn,Carter} .
i) In the first class, atoms are easily trapped by steps as soon as they meet one. 
 The slow process is migration on the terraces. This is migration-limited 
surface-diffusion-kinetics. ii) In the second class, diffusing adatoms
hardly feel steps. The slow process is attachment on steps 
(and detachment from steps). This is  attachment-detachment-limited
surface-diffusion-kinetics. We shall generally consider migration-limited 
surface-diffusion-kinetics. 

Attachment-detachment-limited
surface-diffusion-kinetics is somewhat analogous to 
evaporation-condensation kinetics, since
the adatoms may be considered to have an uniform chemical potential 
$\mu_0$ as if there were a vapour bathing the surface. The exchange of matter 
between this mobile layer and the steps is assumed proportional to 
the difference between $\mu_0$ and the local chemical potential $\mu$  
determined by surface energy changes resulting from surface displacements.

The equilibrium shape of crystals have facets parallel to singular 
orientation~\cite{noz,JVAP}. As seen in the previous section, 
facet-like flat regions also appear in the transient shape of unstable grooves.
A priori, both phenomena are expected to have the same origin.

However, there are significant differences. 
At equilibrium, the chemical potential $\mu$ is the same 
everywhere. The size of a facet  cannot decrease without locally 
increasing the density of adatoms (and therefore the chemical potential) 
near the facet, and the extra adatoms cannot be removed unless they
go back to the facet, which thus comes back to its previous shape.
On the other hand, in the case of the surface perturbations (not at 
equilibrium), 
the chemical potential deviates from its equilibrium value by a positive
amount in the convex regions (top) and by a negative amount in the concave
parts (bottom) and 
the removal of adatoms can be accomplished by transfering them to a
bottom facet. Thus, if facets are for instance artificially  made,
they may be unstable with respect to exchange of atoms from top
facets to bottom facets. Whether this occurs or not is a matter of
dynamics: if the exchange of atoms between facets 
requires too much time, facets can  persist until complete smooting.
This dynamical problem is controversial, as will be seen. 

A simple method~\cite{bonzel3} is to replace the true free energy 
(\ref{phi0}) (actually the improved expression (\ref{phi}) below, which takes 
step interactions into account) by a regularized form (Fig. \ref{fig6}b).
This approach yields facets, and actually the transient groove shape is
not very different from the equilibrium shape~\cite{BW}. However,
the regularization procedure does not give much insight into what really
happens near facets, and it is desirable to devise more microscopic 
methods. Before doing that, it is necessary to discuss interactions
between steps.


\section{Interactions between steps}
\label{ch3}

Formula (\ref{phi0}) consists of just the first two terms of an expansion. 
The next terms depend on interactions between steps. 

Interactions may be short ranged or long ranged. The definition 
which will be accepted in this review is that an interaction
is called `short ranged' if it decays faster than any power $r^{-\alpha}$ 
with the  average distance $r$. Short range interactions  generally 
depend on the microscopic features of each particular material and each 
particular orientation, while long range interactions have 
rather general features which make them especially interesting.


\subsection{Long range interactions}
\label{sch3a}

{\it Electrostatic interaction}
\vskip 0.4cm

The most evident interaction between steps may be the electrostatic one:
symmetry considerations suggest that steps carry an electric dipole 
moment~\cite{thomp}. It is in fact a corollary of the 
electrostatic interaction between adatoms~\cite{klau,lann}. 
Actually, even a flat surface carries an electric moment density: 
it is the correction to this electric moment due to the step that we are 
considering here. 

The interaction between two dipole moments at distance $r$ on the surface of a
solid is proportional to $1/r^3=(x^2+y^2)^{-3/2}$, 
just  as in a homogeneous medium. However, the proportionality factor
is not easy to evaluate. In the case of a metal, different evaluations
have been given  by 
Kohn \& Lau~\cite{klau} and by Lannoo \& Allan~\cite{lann}. Note that
two dipoles at the {\it interface} between two metals would have no
long range interaction because of screening.

If we come back to the case of steps, we can obtain the
interaction energy between  two straight, parallel steps
at a distance $\ell$ by integrating the dipole-dipole interaction 
$(\ell^2+y^2)^{-3/2}$ over $y$. 
The resulting interaction is proportional to $1/\ell^2$, and of course to 
the length of the steps.
The  free energy {\it density} of a regular array of steps at distance
$\ell$ is obtained
 by multiplying by the step density  which is 
itself proportional to $1/\ell=(z_x^2+z_y^2)^{1/2}$ .
The free energy density per unit projected area $\varphi$ (introduced in formula
\ref{8.3}) is then, for $|z_{x}|,|z_{y}| \ll 1$
\begin{equation}
\varphi(z_x,z_y)=\varphi(0,0) + g_1 (z_x^2+z_y^2)^{1/2} + g_3
(z_x^2+z_y^2)^{3/2} 
\label{phi}
\end{equation}
where $g_1=\gamma$ is the step free energy per unit length and 
the constant $g_3=g_3^{elec}$ can be positive or negative, depending of the 
orientation of the dipoles.

\vskip 0.6cm
{\it Elastic interaction}
\vskip 0.4cm

Another type of long range interaction arises from the elastic 
stress which is unavoidably exerted by a step (like any other
defect) on the surface. This stress produces a strain which is of
course maximum near the step, and decreases as $1/r^2$ with the
distance $r$ to this step~\cite{Marchenko,noz,JVAP}.  
As a result, an elastic interaction energy between two parallel steps
at distance $\ell$ arises, which is proportional to  $1/\ell^2$,
just as the electrostatic interaction! For steps of identical sign, 
this interaction is repulsive~\cite{Marchenko,noz,JVAP}. 
When it is taken into account, the resulting 
surface free energy density  of a vicinal surface has still the
form (\ref{phi}), but $g_3$  now contains an
elastic contribution $g_3^{elas}$, which 
is always positive in contrast with $g_3^{elec}$.

A noteworthy exception~\cite{Alerhand} to the $1/\ell^2$ rule is the elastic interaction 
between steps in the case of an adsorbed epitaxial layer which has a 
different lattice parameter from that of the substrate. In this case, the 
(elastic) dipole-dipole interaction has to be integrated over the whole 
misfitting terrace, not only over the step edge: the interaction becomes  
thus logarithmic! 

\vskip 0.6cm
{\it Entropic interaction}
\vskip 0.4cm
 
A third type of interaction between steps arises from the fact that 
steps cannot cross each other. This can be regarded as an
infinite contact repulsion. If the free energy 
density is calculated as a function 
of the average distance $\ell$  between steps, this contact 
repulsion turns out to generate a term proportional to $\ell^{-3}$, just
as in the case of elastic and electrostatic interactions
\cite{Gruber,Pokrovski,Haldane}. 
This term may be understood as resulting from
the reduction of the entropy of a step  by its neighbours. 
Indeed, the entropy is essentially the logarithm of the number of 
degrees of freedom and a step looses degrees of freedom every time it meets
a colleague, since it has no right to cross it. 
This effective interaction may therefore be called {\it entropic}.
It is obviously repulsive since each step tries to have as much space as 
possible. 
 
When the contact repulsion is taken into account, the projected free energy 
density $\varphi$ has still the form (\ref{phi}), but a new term $g_3^{ent}$,
which is positive as  $g_3^{elas}$, has to be added to obtain $g_3$. The
calculations~\cite{PGDG,Gruber,Pokrovski,v80,Haldane,JVAP} of
$g_3^{ent}$ uses the analogy between an array of mutually-avoiding steps 
and the trajectories of non-interaction, one-dimensional fermions. 
The  result~\cite{Haldane} is 
  
\begin{equation}
g_3^{ent} = \frac{\pi^2 (k_BT)^2 }{ 6\tilde\gamma }  \;.
\label{C.13}
\end{equation} 
where $\tilde\gamma$ is the line stiffness.
This result is twice as large as that obtained by Jayaprakash et al~\cite{Jay}  because of a factor 1/2 in the second term of equations (3) and (13) of these authors.
  
  All three interactions of $g_3=g_3^{elec}+g_3^{elas}+g_3^{ent}$ are 
equivalent to a force proportional to $\ell^{-2}$ between pairs of steps,
or to a term  proportional to $\ell^{-3}$ in the free energy 
density. However, there are important differences: the entropic interaction
is a repulsion between nearest neighbour steps only while
elastic and electrostatic interactions relate all pairs of neighbours.

\vskip 0.6cm
{\it Experiments}
\vskip 0.4cm
  
The case of the solid-superfluid interface of helium is 
particularly accessible to experiments 
because equilibrium is much more rapidly established than in other 
materials. Certain experiments~\cite{Andreeva} suggest an interaction 
which decreases  slower than $r^{-2}$. A similar suggestion has
been made by Arenhold et al~\cite{aren} for Pb. 
A theoretical attempt to explain that have been made
(in the case of He) by Uwaha~\cite{uwa}. On the other hand, 
Balibar et al~\cite{Balibar} have pointed out that experimental results must be 
very cautiously interpreted.

  
\subsection{Oscillating and short range interactions}
\label{sch3b}

As said before, short range interactions are generally particular to 
each material and it is difficult to say anything general.
Oscillating interactions have similar effects since they have local 
minima at short distances. 
One can mention the following  mechanisms.
  
\vskip 0.6cm
{\it Oscillating electronic interactions}
\vskip 0.4cm
  
For certain metal surfaces
there is a partially unfilled electronic surface band. 
Then a phenomenon analogous to Friedel oscillations generates 
an oscillating interaction~\cite{lau78} between point defects.
This interaction is proportional to $r^{-2}\cos(2k_F r)$, 
where $k_F$ is the Fermi wavevector and $r$ the distance between defects. 
Integrating along 
a  row generates an interaction between parallel line defects
(e.g. steps) which is proportional  to $r^{-1}\cos(2k_F r)$. 
 Since it oscillates  
with distance, the absolute minimum 
corresponds to a finite value of the distance $\ell_0$  between steps. 
The  effect is very important since  surface orientations with steps larger 
than $\ell_0$ 
can become unstable as observed experimentally in certain cases 
(see subsection \ref{sch2}). 
Oscillations are washed out when 
 $r$ becomes longer than the electron mean free path, and therefore the  
interactions between parallel steps on a clean surface are likely to decay 
as $1/r^2$ at long distances in all cases. An oscillating interaction  
appears also without partially unfilled electronic surface band, 
but it decreases faster with distance~\cite{lau78}.

\vskip 0.6cm
{\it Interactions resulting from reconstruction}
\vskip 0.4cm
  
Surface reconstruction may have a similar effect, i.e. make certain
orientations  more favourable than the intermediate one, in other words, 
create an oscillating interaction between steps.~\cite{Liu} 
  
A typical example of reconstruction is provided by the (001) surface of 
silicon or any semiconductor. Semiconductors usually have the diamond
structure, in which atoms have a very low coordination (4 neighbours). 
Creating a (001) face in a rigid lattice would reduce  the coordination 
of surface atoms to two neighbours and that would cost a lot of  energy. 
Therefore the surface atoms form `dimers', thus reducing the
symmetry. This symmetry reduction with respect to that of the rigid  
lattice is called {\it surface reconstruction}.  Now, step positions  
which would cut dimers are unfavourable, and this disavantages certain
distances  between steps.

The  orientations which are favoured by reconstruction are 
sometimes called `magic'~\cite{Bartolini}.


\section{Mean field theory for a well-cut singular surface}
\label{ch4}

 After having recalled the various general properties of surfaces and steps, 
 we come back to the experiments on groove smoothing 
on Ni(001) and Ni(111) described in subsection
\ref{sch2.1}. The case of Au(001) and Au(111), whose vicinals are unstable,
is excluded here and will be addressed in section \ref{ch11}

Steps are complicated objects and in this section we shall 
 yield to the temptation to consider them as straight lines 
(Fig. \ref{rettovi} a).
 This is only possible if the average surface is initially 
 perfectly cut along a symmetry orientation and if there are no dislocations.
 Even then, steps are not straight, but fluctuating lines 
 (Fig. \ref{rettovi} b).
 Neglecting these fluctuations may be called a ``mean field 
 approximation''. This approach was chosen by Rettori \& Villain~\cite{Rettori}
 and Ozdemir \& Zangwill~\cite{Ozdemir}, who assumed an interaction energy 
proportional to 
 $L/\ell^2$ between nearest neighbour steps of length $L$ at distance $\ell$.
 These features are those of the entropic 
 interaction addressed in subsection \ref{sch3a}. 

We will  recall here the  results of  Rettori and Villain~\cite{Rettori} 
and Ozdemir and Zangwill~\cite{Ozdemir}.
Below $T_R$, the analytic form (\ref{8.4}) of the projected free energy 
$\varphi$ is not correct. Instead, one should use (\ref{phi}), where 
$g_3$ is given by (\ref{C.13}) in the absence of step interactions
except  contact interaction.
In a continuum approximation, the instantaneous shape of the
surface can be described by a height function  $ z(x,t)$ 
(if the perturbation is unidirectional) and (\ref{phi}) takes the form 
\begin{equation}
\varphi=g_1|z_x|+g_3|z_x|^3
\label{gruber}
\end{equation}

 Using (\ref{8.3}) and
(\ref{gruber}), one  obtains the excess
 chemical potential as~\cite{Rettori,Ozdemir}
\begin{equation}
\delta \mu=-6a^2g_3|z_x|z_{xx}
\label{potvi}
\end{equation}
where  $z_{xx}=\partial^2 z/\partial x^2$. 
 
If the relaxation occurs by surface diffusion, we can still write 
(\ref{diffusion}) and (\ref{current}),  and together with  (\ref{potvi}) 
the equation of motion is obtained
\begin{equation}
\frac{\partial z}{\partial t} = -G_3 
\frac {\partial^2}{\partial x^2}|z_x|z_{xx}
\label{motion}
\end{equation}
where $G_3 \propto g_3 \tilde D_s / k_B T$ and $\tilde D_s = a^2 \rho_0 D  $
as before.
Eq.(\ref{motion}) is  non-linear and  until now no analytic 
solution has been found for it. 

Since $\delta \mu$ should vary smoothly in space, (\ref{potvi}) is
consistent~\cite{Lancon} with a singularity near the top 
($\delta x \simeq 0$) of the type 

\begin{equation}
z=h(t)-C(t)|\delta x|^{3/2} 
\label{eq3}
\end{equation}
where $h(t)$ is the amplitude of the profile at time $t$.

The singularity (\ref{eq3}) is the same as predicted at equilibrium near a 
facet. However, according to  Rettori \& Villain and to 
Ozdenir \& Zangwill no facets are expected to form during flattening. 
Indeed, the decay of the top (and bottom) terraces 
(should they form facets or not) is due to the junction of the top (or 
bottom, respectively) ledges. When two top (or bottom) ledges approach from each
other, why should lower ledges stay behind instead of 
keeping contact with them?
If top ledges keep contact with lower ledges, 
no facet is expected in the decaying profile. Therefore, according to 
(\ref{eq3}), 
the profile is predicted to be sharper than a parabola at its top 
(and at its bottom as well). 
   
Ozdemir \& Zangwill have solved (\ref{motion}), or
more precisely a discrete version of (\ref{motion}),
and found under certain assumptions (shape preserving  solution) 
that the amplitude $h(t)$ decays with time $t$ as 

\begin{equation}
h(t)=h(0)\frac{1}{1+t/\tau_\lambda}
\label{hhhh}
\end{equation}
where $\tau_\lambda \simeq \lambda^5$.
Relation (\ref{hhhh}) is often quoted~\cite{Jiang1,Murty} but it is not always 
mentioned that $\tau_\lambda$ has to depend on $h(0)$. Indeed, if (\ref{hhhh})
holds for a particular value $h_0$ of $h(0)$, with a particular value
$\tau_\lambda^0$ of $\tau_\lambda$, it follows, for any time $t_1$
($0<t_1<t$),
$$
h(t)=h(0)\frac{1}{1+t_1/\tau_\lambda^0} 
\frac{1+t_1/\tau_\lambda^0}{1+t/\tau_\lambda^0}
=h(t_1) \frac{1+t_1/\tau_\lambda^0}{1+t/\tau_\lambda^0}
=\frac{h(t_1) }{1+(t-t_1)/(t_1+\tau_\lambda^0)}
=\frac{h(t_1) }{1+h(t_1)(t-t_1)/(h_0 \tau_\lambda^0)}
$$

If the time origin is translated to $t_1$, this coincides with (\ref{hhhh}),
but $\tau_\lambda$ is proportional to $1/h(0)$:

\begin{equation}
\tau_\lambda = \tau_\lambda^0 h_0/h(0)
\label{hhhhh}
\end{equation}

The weakness of the above theory is that any interaction between
steps of  opposite sign (at the top and the bottom of the profile)
is ignored. This would be acceptable if the ledges were straight
(Fig. \ref{rettovi} a). In reality, they have fluctuations
(Fig. \ref{rettovi} b). We shall try to take these fluctuations into 
account in section \ref{ch8}.


\section{Facet-predicting theories or the ambiguity of the chemical potential}
\label{ch5}

A paradox of the mean field theory outlined in the previous section
is  the  energy release which results from the pairwise annihilation of 
ledges at the top and bottom of the profile  when they are assumed to be straight. 
Spohn~\cite{Spohn} and Hager \&  Spohn~\cite{Hager} have proposed a completely different theory of groove smoothing which 
does not show this paradox, at least in the case of surface diffusion kinetics. 
In the case of evaporation-condensation kinetics, both approaches 
predicts the  same singularity, a sharpening 
of the profile at maxima and minima for long times : 
$z \sim {| \delta x |}^{4/3}  $ (this prediction is only valid in the limit 
of a slowly varying surface profile, i.e., a minimum sample size is needed). 

The remainder of this section is devoted to surface diffusion kinetics.
In this case, Spohn's results regarding the shape 
of the profile are in strong contrast with 
those described in section \ref{ch4}. He finds, indeed, that facets form 
in the transient regime.

As pointed out by Tang~\cite{Tang}, the reason of the disagreement lies in  
different expressions of the local chemical potential $\mu$ on  
a surface which is not at equilibrium. Since the current is a function of $\nabla \mu$ 
as seen above, the consequences are crucial. 

The  chemical potential $\mu$ is  the sum of
its value $\mu_0$ in the solid far from the surface, and 
an {\it excess} $\delta \mu$ resulting from the surface tension.

If a system has $N$ atoms and a free energy $F$, its chemical potential
is  $\mu=F/N$. 

Now, let us consider the topmost terrace of the profile of Figure 
\ref{rettovi} a.
If its width is $\ell$ and if the sample length is $L$, then 
the number of atoms in the terrace is $N=\ell L/ a^2$
while, if interactions between steps are neglected, 
the free energy excess resulting from the surface energy is 
$\delta F=2\gamma L$ where $\gamma$ is the step free
energy per unit length. If the whole terrace is removed, the 
corresponding variation of free energy is $N \mu_0 + \delta F$
hence $\delta \mu = \delta F/N$ or
\begin{equation}
\delta \mu = 2\gamma a^2/\ell
\label{spohn}
\end{equation}

Experiments are performed on macroscopic samples, where this
expression is much larger than the term coming from step interactions.

If (\ref{spohn}) is applied to  grooves, the 
same formula holds for the bottom terrace with an opposite sign, 
$\delta \mu = -2\gamma a^2/\ell$ if the bottom terrace has the same width.
The average chemical potential gradient is therefore  
$ \nabla\mu = 4 \gamma a^2/(\ell\lambda) $ and the current
from the top to the bottom is of order $ 4 D \gamma a^2/(\ell\lambda)$. 
If $\ell$ is small, the bottom terraces are rapidly filled and the top 
terraces are rapidly peeled, and after some time, only the intermediate 
terraces are left, and facets appear at the top and at the bottom. 
This is Spohn's description. 

In section \ref{ch4}, on the other hand, an underlying idea is the
following: if the edges of the topmost terrace move so that the width changes
by {\it small} amount $d \ell$, the number of atoms of that
terrace changes by $d N=2Ld \ell/a^2$, but 
the total length of the   edges does not change,
it is always $2L$, and therefore the chemical potential is,
{\it disregarding interactions} between steps, $\delta \mu= 0/d N$ or 

\begin{equation}
\delta \mu = 0
\label{zero}
\end{equation}

Instead of a straight step, it is of interest to consider also
a closed, e.g. circular terrace of size $R$. It contains $N=\pi R^2/a^2$,
where $a^2$ is the atomic area, and its free energy excess is
$\delta F = 2\pi \gamma R$. The formula analogous to (\ref{spohn}) is

\begin{equation}
\delta \mu = \delta F /N = 2\gamma a^2/R
\label{spohn'}
\end{equation}
while, arguing along the lines of section \ref{ch4}, one
obtains~\cite{Lancon}  
\begin{equation}
\delta \mu = d(\delta F)/dN = \frac{\gamma a^2}{R} 
\label{GibbsT}
\end{equation}

Formulae (\ref{spohn'}) and (\ref{GibbsT}) are still different, but just
by a factor 2. The results which can be deduced from both formulae are
just quantitatively different, not qualitatively. 
In any case smaller terraces 
have the higher chemical potential and therefore decay first.

Formulae  (\ref{spohn'}) and (\ref{GibbsT}) are in contradiction.
Which one  is correct at equilibrium? It depends.

For a single terrace on a high symmetry surface, (\ref{GibbsT}) is correct
for a circular terrace of radius $R$, while $\delta \mu=0$
for a stripe-shaped terrace, as can be obtained 
from (\ref{GibbsT}) by replacing the ledge radius $R$ by $\infty$.

However, if a finite crystal has an approximately circular facet
of radius $R$ at equilibrium, $\delta \mu$ can be obtained by
removing a whole atomic layer from a facet. The step interaction energy is
not much modified in that operation, and therefore  (\ref{spohn'}).
If the crystal is obliged to have the shape of a bar, $\delta \mu$
is given at equilibrium by (\ref{spohn}). Of course, it is also given by
(\ref{potvi}), and both expressions should be equal at equilibrium.

Out of equilibrium, there is no reason {\it a priori} to write 
(\ref{spohn}) or (\ref{zero}), (\ref{spohn'}) or (\ref{GibbsT}). A careful 
analysis of the dynamics is necessary to solve the controversy.

Relation  (\ref{GibbsT}), which relates the chemical potential
to the curvature, is a two-dimensional form 
of the Gibbs-Thomson formula. 

An objection to Spohn's theory is that it does not describe how
pairs of top (or bottom) steps merge. The peeling  of extreme layers
are accompanied by fluctuations which are not better taken into account by him
than by Rettori \& Villain
and Ozdemir \& Zangwill. In the next section we shall try to find 
a solution of the controversy in simulations.


\section{Simulations}
\label{ch6}
The relaxation of profiles have been simulated in (1+1)D and (2+1)D by many authors
\cite{Selke,Jiang,Jiang1,Erle,Murty,Searson,Selke1,Tang,Erwan}. 
The  solid-on-solid (SOS) model have been used by most of them. Its Hamiltonian is 
\begin{equation}
{\cal H} = J \sum_{xy,x'y'} |h_{xy}-h_{x' y'}|
\end{equation}
where $h_{xy}$ denotes the height  of the surface at site $xy$, the sum runs 
over nearest-neighbours pairs of sites and $J$ is the bonding energy.

Both  diffusion and  evaporation dynamics were considered in the simulations,
and in general the standard Metropolis algorith was used.

For evaporation-condensation,  a conserved dynamics is sometimes used, in the 
form of a detailed balance of particle exchange between the surface and the 
surrounding gas: a surface atom can move from its initial position to some 
other, arbitrarily distant site (conserved number of surface atoms). In some 
works, non-conserved dynamics are used.
For (2D+1) and  $T>T_R$, the simulation results agree well with the 
predictions of Mullins: the profile relaxes as a sinusoid~\cite{Oitmaa,Selke0}. 
For $T<T_R$ it was observed~\cite{Selke0} that the profile fluctuates, 
in time, between a sinusoidal and a ``trapezoidal'' form. 
Selke and Bieker~\cite{Selke0} observed that these fluctuations are 
accompanied by two distinct time scales in the flattening procedure, 
reflecting the step wandering and the shrinking of islands. 
They assumed that the  meandering stage leads to the dominant time scale 
and their results for  the relaxation time $\tau$ are
 compatible with  $\tau \propto \lambda^3$, but they speculate that 
for larger systems sizes, $\tau \propto \lambda^4$. Actually, recent large scale simulations  by Tang~\cite{Tang} 
confirmed  the exponent $4$, but with a  correction in the power law.

For diffusion,  the number of particles  must be conserved and 
particle exchange  occur between nearest-neighbors sites.
For temperatures above $T_R$ the results of the 
simulations~\cite{Selke0,Searson,Murty} agree well with the predictions of 
Mullins (sinusoidal profile decay~\cite{Selke0}, the value  $n=4$ was obtained 
for the scaling law $\tau \propto \lambda^n$ of the relaxation time~\cite{Searson,Murty}). 
For temperatures  $T < T_R$, some authors~\cite{Jiang1} found that 
the scaling behaviour breaks down after some time and then cusps and 
plateaus develop at the top and bottom of the profile, 
others~\cite{Selke0,Erle} found that the shape of the relaxation 
profile fluctuates in time between having a flat top to having a rounded 
top (sinusoidal). We remark that these authors 
have considered typically small initial profile heights  $h=4,5$.
Different scaling laws  for the typical relaxation time $\tau$ were obtained: 
$n=4$ (obtained by Searson et al~\cite{Searson}), 
$n=3.5 - 4 $ (obtained by Erlebacher et al~\cite{Erle}),  
$n=5$ (obtained by Ramana Murty et al~\cite{Murty}).  

We mention that  most authors performed their simulations using  standard 
Monte Carlo algorithms, and from that they obtained time dependent laws for 
the quantities of interest. Presently it becomes more clear~\cite{Erle,Kotrla} 
that 
the kinetics of the relaxation must be preferable studied by methods such as 
the Kinetic Monte Carlo (KMC), which includes a prescription  to take into 
account the real time instead of to take the unit of time as the interval 
between two updatings, for instance. Only very recently KMC simulations have 
been performed\cite{Erle,Murty,Erwan} on the relaxation of grooves.

Another important point is that, since in the evaporation-condensation 
dynamics atoms are exchanged between the surface and the vapour and  
in diffusion dynamics it is necessary to consider the motion of the atoms 
from a site to another (in general a  nearest-neighbour),  surface diffusion  
takes  much more computing time than evaporation-condensation. Due to that, 
in many simulations (for diffusion dynamics) some short-circuit artefact have 
been introduced, especially,  small  samples have been considered (small 
heights, or small widths).

In what concern the choice  of small samples, one of the problems is that  
to obtain the correct scaling laws
(e.g. the relaxation time $\tau$ vs the wavelength $\lambda$),
the amplitude $h$ and the average distance between steps
$\ell=a\lambda/(4h)$ must be large with respect to the atomic distance $a$. 

{\it What is a short sample?}

The problem related to the width $L$ of the system in the direction of
the grooves is different. In available analytical or numerical theories, 
elastic or electrostatic interactions between the steps which form the 
profile are not considered, and then the only interaction 
which remains is the contact repulsion.
As seen in section \ref{sch3a}, in analytic theories~\cite{Rettori,Ozdemir,Spohn,Hager}
this contact repulsion is taken into account 
by introducing an effective free energy equal
~\cite{Gruber,Pokrovski,Haldane} to
\begin{equation}
f_{ent}= L\frac{(\pi k_BT)^2}{6\tilde\gamma \ell^2}
\label{talapov}
\end{equation}
for each pair of steps at average distance $\ell$.

Here, $\tilde\gamma$ is the line stiffness related to the energy $W$
per kink by the relation
\begin{equation}
\beta\tilde\gamma a= 1 + \frac{\exp(\beta W)}{2}
\label{gamma}
\end{equation}

Since  steps are not straight at finite temperatures, it is useful 
to define the average distance $\xi$ between kinks along a step,  
given by
\begin{equation}
\xi \approx a \left( 1+\frac{\exp(\beta W)}{2} \right).
\label{xi}
\end{equation}

Formula~(\ref{talapov}) is correct only for long distances $\ell$.
Moreover (\ref{talapov}) makes sense only for large $L$.
An exact calculation\cite{v80}
--for instance considering a single fluctuating line between 
two straight lines~\cite{ff} at distance $2\ell$--
shows that (\ref{talapov}) holds only if the relation  
\begin{equation}
L \gg   \xi \ell^2/a^2
\label{ineq}
\end{equation}
is satisfied. This yields a first condition on $\xi$
\begin{equation}
\xi \ll L (4h/ \lambda)^2 .
\label{ineq'}
\end{equation}
Whether this condition is satisfied or not depends on the temperature T,
which should anyway be smaller than $T_R$.
An order of magnitude of $T_R$ is given by
writing that the free energy of an isolated step per atom, which is 
$W - k_BT \ln[1+2\exp(-\beta W)]$ at low temperature, vanishes. Thus
\begin{equation}
\exp\frac{W}{k_BT_R} \simeq 2
\label{TR}
\end{equation}
 
As a matter of fact, if one wishes to check
the analytic predictions made for singular surfaces,
it is safer to choose $T$ much lower than $T_R$. Indeed the 
aspect of a surface on short distances is very similar just below $T_R$
and just above $T_R$. In particular, closed terraces are present between 
steps, and steps exhibit ``overhangs'' which are generally ignored in
available theoretical treatments. The choice $T<T_R/2$ seems reasonable. 
This implies, according to (\ref{xi}) and (\ref{TR}), a second condition on $\xi$
\begin{equation}
\label{eq4'}
\xi > 3a
\label{ineg}
\end{equation}

It turns out that conditions (\ref{ineq'}) and (\ref{eq4'})
are not always simultaneously satisfied in simulations, 
specially when diffusion kinetics is considered~\cite{Jiang1,Erle}.
The simulation by Ramana Murty and Cooper~\cite{Murty} is a special case, 
since they  used sufficient long samples, and actually 
Ramana Murty and Cooper do find the scaling law $\tau \propto \lambda^5$ predicted
by Ozdemir and Zangwill. Note, however, the very low value of $h$ in all
simulations. The choice of all lengths ($\lambda$, $h$, $L$) 
results from a compromise between too large values which would 
saturate computers, and too short sizes which would have no relation with 
reality.

A conclusion which can be drawn from  simulations~\cite{Erwan}
is that facets do appear at least in certain cases,
but it is not quite sure that this phenomenon is not related to
the small sample size. For a given sample length, facets
seem to be observed in simulations at low temperature, when the 
steps have but few contact points. On the other hand, at higher
temperatures, the profile is roughly sinusoidal~\cite{Erwan'}.
In the following section, an explanation of the results 
on short samples is presented.

\section{The transient attraction}
\label{ch7}

The interactions between steps we considered so far are thermodynamic.
They correspond to terms of the free energy, and have an effect on the 
equilibrium state. When writing kinetic equations, these thermodynamic 
interactions appear in the detailed balance relation

\begin{equation}
p_A^B \exp(-\beta E_A) = p_B^A \exp(-\beta E_B)
\label{db}
\end{equation}
where $p_A^B$ denotes
transition probability per unit time  from a state $A$ to state $B$
and $E_A$ is the energy of state $A$.

In a non-equilibrium situation, we will prove that 
a clustering  of steps of identical sign can occur in the case of 
surface diffusion kinetics. 
This can be interpreted as arising from an attraction 
between such steps, even though there is no thermodynamic interaction, i.e.
$E_A=E_B$ has the same value for all configurations of interest. 

Unfortunately, a quantitative theory of this attraction can only be 
done for a 1-dimensional surface (Fig. \ref{paires} a)
which is a good representation of a short sample.~\cite{Erwan} 

The one-dimensional model is fully characterized 
by the probabilities per unit time 
and unit length, $\alpha^+(\ell)$ and $\alpha^-(\ell)$,
for two neighbouring steps at distance $\ell$, to exchange
an atom in one direction or in the other one. 
The case of interest is when both steps have identical sign.
Consider the configuration $B$ obtained from a configuration $A$ by
transfering one atom downward from a step to the neighbouring one
at distance $\ell$.

The transition probability per unit time from $A$ to $B$ is, by definition,
$\alpha^+(\ell)$. The transition probability per unit time from $B$ to $A$
is $\alpha^-(\ell+2a)$ because the terrace width in $B$ is $\ell+2a$. 
In the absence of interaction between steps,
the detailed balance relation writes
\begin{equation}
\alpha^-(\ell+2a) = \alpha^+(\ell)
\label{bilan}
\end{equation}
 
If the distance $\ell$ between both steps is large,
most of the emitted atoms go back to the step 
where they started from. The transfer probabilities $\alpha^-(\ell)$
and  $\alpha^+(\ell)$ are therefore expected to decrease with
increasing $\ell$.

More precisely, it can be shown~\cite{Erwan}, that:
\begin{equation}
\alpha^+(\ell)=\alpha_0 \frac{a}{\ell+\ell_s+a}
\label{alfa+}
\end{equation}
and
\begin{equation}
\alpha^-(\ell)=\alpha_0 \frac{a}{\ell+\ell_s-a}
\label{alfa-}
\end{equation}
where $\alpha_0$ is a constant and the ``Schwoebel 
length''~\cite{Ehrlich,Schwoebel} $\ell_s$
(equal to 0 in most of available simulations) 
depends on the detachment probability of atoms from steps.
The emission rate $\alpha_0a$ per lattice site
is obtained if one notices that, at equilibrium,
adatom emission by steps is compensated by absorption. The absorption rate
is proportional to the product of the adatom density $\rho_0$  by the adatom 
jumping rate $4 D/a^2$, where $D$ is the adatom diffusion constant. Therefore
\begin{equation}
\alpha_0=\frac{4D \rho_0}{a} 
\label{KKK}\;\;.
\end{equation}

There is no thermodynamic interaction between steps at equilibrium since 
(\ref{alfa+}) and (\ref{alfa-}) satisfy the detailed balance
relation (\ref{bilan}). However, $\alpha^-(\ell)>\alpha^+(\ell)$,
which means that atom exchange between two steps of
identical sign tends to shorten their  distance. 
It is hardly avoidable to interpret this  as a kind 
of attraction between neighbouring steps of identical sign.

Of course, all steps should be taken into account, and the effect disappears 
for a regular array of equidistant 
steps (an equilibrium configuration) but it is present for the
periodic profile of Fig. \ref{paires} a, as demonstrated by Monte-Carlo 
simulations~\cite{Erwan}. The result is a blunting of the profile;
steps of identical sign attract themselves, so that the maximal slope
increases, and consequently the top and the bottom flatten.

The following exercise helps to understand the nature of the 
interaction. Consider a one-dimensional profile made of alternating 
pairs of up and down steps (Fig \ref{paires} b).
Let the distance between 
two consecutive steps be initially uniform and equal to $\ell_0$.
It will first be assumed that steps of opposite sign cannot annihilate
--an unphysical model which has the advantage to have a non-trivial 
equilibrium state in which the average distance between two consecutive steps
is $\ell_0$. In this model
the average distance $<\ell(t)>$ betwen two consecutive steps
of identical sign will first decrease because of the attractive interaction,
until collisions take place. Then, as an effect of the contact repulsion, 
$<\ell(t)>$ will increase and come back to its equilibrium value $\ell_0$.
Thus, the attraction has just a transient effect. For this reason, we will call 
it {\it transient attraction}.

In a one-dimensional profile with alternating groups of $2h/c$ up steps and
 $2h/c$ down steps, the transient attraction does produce an 
initial clustering of 
identical steps, which produces facets, which are observed in
Monte-Carlo simulations~\cite{Erwan}. In these simulations, steps of
opposite sign are allowed to annihilate, and it turns out that facets persist
until the profile completely flattens. This is a surprise, 
but the effect seems to be observed only on one-dimensional
samples or in two-dimensional ones which are too short in
the groove direction. 
Therefore, the transient attraction will be ignored in the following sections.


\section{Step fluctuations at the top and bottom}
\label{ch8}

Analytic theories of  Rettori and Villain~\cite{Rettori} and
Ozdemir and Zangwill~\cite{Ozdemir} are  mean field ones, where fluctuations 
are neglected.  Away from the top and bottom, step fluctuations are expected to
be reasonably taken into account by the entropic interaction. 
The challenge is now to introduce  step fluctuations of the
top and bottom ledges into these theories. The following 
calculation~\cite{mystere} is an attempt to solve this problem.

Because of these fluctuations, 
finite terraces, i.e. closed step loops, form  at the top of the profile,
which have a typical length $L_1$ (Fig. \ref{rettovi}b). 
The width of the loop should be comparable 
with the distance $2\ell_1$ between both topmost infinite
steps of each period.
Each finite terrace terminates by a tip whose radius 
is of order $\ell_1$, so that the excess chemical 
potential before the tip is 

\begin{equation}
\delta \mu=\frac{\gamma a^2}{\ell_1} 
\label{7bis}
\end{equation}

 The velocity $v$
of the tip is proportional to  the current 
density of adatoms flowing from the tip which is equal  
to $ \beta \rho_0  D \vec \nabla \mu$.
Since the tip is at distance $\ell_1$ of the terrace edges,
the order of magnitude of $\nabla \mu$ should be expression
(\ref{7bis}) divided by  $\ell_1$, so that

\begin{equation}
v \approx \frac{\beta \rho_0 D \gamma a^4}{\ell_1^2}
\label{eq9}
\end{equation}

The time $\tau_1$ necessary to peel the upper layer 
is obviously 
 
 \begin{equation}
\tau_1=L_1/v
\label{eq9'}
\end{equation}
so that (\ref{eq9}) yields

\begin{equation}
\tau_1 \approx \frac{L_1\ell_1^2}{\beta \rho_0  D \gamma a^4}
\label{2.1}
\end{equation}

The nucleation time of a finite terrace after the
previous one has disappeared is  expected~\cite{Duport}
to have the same order of magnitude
$\tau$ as the lifetime of a finite terrace.

Topmost steps can only fluctuate by emission of atoms. These atoms can
only go to the lowest steps. This 
suggests that 
step fluctuations  arise  from the exchange of atoms
between  topmost and  lowest terraces. 
This exchange is of course biased,
and preferably from the top to the bottom. But it is also partly stochastic.
Its mean square fluctuation will be calculated without taking the bias 
into account. The exchange 
rate in each direction is expected to be given by a formula 
analogous to (\ref{alfa+})
or (\ref{alfa-}) where the distance $\ell$ is replaced by the half-wavelength 
$\lambda/2$ while the proportionality factor is given by 
(\ref{KKK}). Thus, the  average number $d \nu/(dt dL)$ 
of atoms exchanged from
the top to the bottom  per unit 
length in the unit time is (neglecting bias)
 \begin{equation}
 \frac{d \nu}{dt dL} \approx \frac{8 \rho_0 D}{\lambda}
\label{5.1b}
\end{equation}

The average $<\nu>$ of the number $\nu$ of particles emitted
 by a length $b$ of a topmost step in
the time $\tau_1$ is the product of this quantity by $b\tau_1$. The
 mean square fluctuation $<\delta \nu^2>$ of this number  is, if the various
 absorption and emission processes are independent:
  
 \begin{equation}
<\delta \nu^2>=<\nu>=\frac{d \nu}{d t d L}b \tau_1
\label{soupe}
\end{equation}

This fluctuation of the number of atoms produces a fluctuation
$\delta \ell$ of the distance between the topmost steps on a length $b$. 
Expression (\ref{soupe}) can therefore be identified with the square of the 
area $b\delta \ell$, so that

 \begin{equation}
<\delta \ell^2>=\frac{a^4 <\nu>}{b^2}=\frac{d \nu}{d t d L} \frac{\tau_1 a^4}{b}
\label{potage}
\end{equation}
 
The length
$b$ should be chosen as the most efficient one, which maximises (\ref{potage}). 
Therefore, $b$ should be chosen as short as possible. However, expression
(\ref{potage}) cannot be larger than the thermal fluctuation, so that 
$<\delta \ell^2><k_B T b/\gamma$. The appropriate choice of $b$
is therefore

\begin{equation}
b = \frac{\gamma <\delta \ell^2>}{k_B T }
\label{brouet}
\end{equation}
 
The topmost steps meet when $<\delta \ell^2>=\ell^2$. This relation,
together with (\ref{potage}), (\ref{2.1}), (\ref{brouet}) and (\ref{5.1b}),
yields an expression for $L_1$:

\begin{equation}
 L_1 \approx \frac{ \gamma^2 \lambda \ell^2}{8k_B^2T^2}
\label{5.3}
\end{equation}

The average current density $j$ from a top terrace is obtained 
by calculating the total current
from each tip, which is proportional to the product of  (\ref{eq9}) by
$\ell$, and dividing by the distance $L_1$ 

\begin{equation}
j \approx \frac{\rho_0 \beta D \gamma a^2}{L_1\ell_1}
\label{3.2}
\end{equation}

This can be identified with  the current density of atoms from the
top to the bottom terrace, which can also be roughly evaluated  by  dividing 
the chemical potential difference $2\delta\mu$
(given by  formulae  \ref{potvi} and \ref{eq3})
by  $\lambda/2$ (in order to
obtain the chemical potential gradient)   and then
multiplying by the same kinetic coefficient 
$\rho_0 \beta D$ as in (\ref{5.1b}).

 \begin{equation} 
j =\frac{27C^2\rho_0 \beta Dg_3 a^2}{2\lambda}
\label{3.1}
\end{equation}

Identification of (\ref{3.2}) and (\ref{3.1}) yields

\begin{equation} 
L_1 = \frac{2\lambda\gamma}{27C^2   g_3 \ell_1}
\label{LLL}
\end{equation}

Eliminating $L_1$ between (\ref{LLL}) and (\ref{5.3}), one obtains

 \begin{equation} 
\ell_1^3 \approx \frac{16(k_BT)^2}{27C^2g_3 \gamma }
\label{enfin}
\end{equation}

If $g_3$ is replaced by its value (\ref{C.13}),  this relation reduces to
$\ell^3 \approx C^{-2}$. This is just what could be deduced from
(\ref{eq3}) by replacing $h-z$ by 1 and $\delta x$ by $\ell_1$. In other words, 
the results of the mean field theory of  section \ref{ch4} would
not be considerably modified if fluctuations were taken into account. 

An approximate expression of $C(t)$ as a function of the modulation amplitude
$h(t)$ can be obtained if one assumes that (\ref{eq3}) is approximately 
correct for $z=0$.  Then, since $\lambda\gg \ell_1$, 

\begin{equation}
C(t) \approx h(t)\left(\lambda/4 \right)^{-3/2} 
\label{eq7}
\end{equation}

The peeling time of the top layers is given by inserting
(\ref{LLL}) and (\ref{enfin}) into (\ref{2.1}). One obtains

$$
\tau_1 \approx \frac{ \gamma^2 \lambda }{8k_B^2T^2}
\frac{\ell_1^4}{\beta \rho_0  D \gamma a^4}
 \approx \frac{ \gamma^2 \lambda }{8k_B^2T^2}
\frac{1}{\beta \rho_0  D \gamma a^4}
\left[\frac{16(k_BT)^2}{27g_3 \gamma }\right]^{4/3}C^{-8/3}
$$
and, using (\ref{eq7}),
\begin{equation}
\tau_1 \approx \frac{ \gamma^2 }{2k_B^2T^2}
\frac{1}{\beta \rho_0  D \gamma a^4}
\left[\frac{16(k_BT)^2}{27g_3 \gamma }\right]^{4/3}h(t)^{-8/3}
 \left(\frac{\lambda}{4} \right)^{5} 
\label{2.111}
\end{equation}

The quantity which can easily be measured is the time after which
the amplitude $h$ has been reduced by a factor 2 for instance. 
This decay time is

\begin{equation}
\tau_{dec} \approx \tau_1 h \simeq h^{-5/3} \lambda^{5} \;\;.
\label{2.112}
\end{equation}

The power $\lambda^{5}$ is the same as obtained by Ozdemir and Zangwill
(see formula \ref{hhhhh}) but expression (\ref{hhhhh}) was proportional 
to $1/h$.
The amplitudes $h$ used in Monte-Carlo simulations are too small to
decide which formula is correct, and so far as we know, there is no 
experiment which can prove or disprove (\ref{2.112}).


\section{Miscut surfaces}
\label{ch9}

Metal surfaces cannot easily be cut along a high-symmetry direction. There 
is generally a small miscut angle $\alpha$. If $x$ is the direction common to the high 
symmetry plane 
and to the miscut average surface, the groove direction is defined by its angle
$(\pi/2+\psi)$ with $x$. A sinusoidal surface modulation 
causes the steps to assume a specific shape. For $\psi =0$ they are sinusoidal 
in an projected on-top view while for $\psi=\pi/2$ they are straight, with 
their 
separation modulated sinusoidally. Staying with a low amplitude/wavelength 
ratio (surface slopes small compared to $\alpha$) and using a surface free 
energy  of the kind in eq.(\ref{phi}) solutions for the rate of profile decay 
via migration-limited, surface diffusion kinetics 
were derived  by Bonzel and Mullins~\cite{Bonzel2} for two particular 
cases: $\psi=0$
and $\psi=\pi/2$. In the first case the rate is proportional to
$(g_1 + 3g_3 \tan^2\alpha)/\sin \alpha$, which is approximately equal to 
$g_1/\sin\alpha$ 
because $g_3 \tan^2\alpha \ll g_1$. Thus the step self-energy 
is the important energetic quantity, driving the profile decay. This is 
understandable because the wavy steps want to reduce energy by becoming 
straight which is equivalent to a reduction in profile amplitude. For  
$\psi=\pi/2$, on the other hand, the decay rate is proportional to
$6g_3\sin\alpha/\cos^4\alpha$, 
i.e. it is driven by the step interaction energy only. Since the steps are 
basically straight from the beginning, the total energy can only be reduced 
by a redistribution of all steps, so that their separation becomes about 
equal. This redistribution of steps causes the amplitude 
to decrease. The ratio of rates for these two types of modulation 
is approximately $g_1/(6g_3\sin^2\alpha)$ for 
$\alpha < 10^\circ$ assuming equal surface diffusion coefficients normal 
and parallel to 
steps. Since $g_1$ is believed to be larger than $g_3$, this ratio can be a 
large 
number, depending on the choice of $\alpha$. 
An experimental verification of the 
effect was given for 1.5$^\circ$ and 5$^\circ$ miscut Au(111) surfaces that 
had been 
modulated in the $\psi=0$ and $\psi=\pi/2$ modes.
Figure \ref{xxx} 
shows the experimental amplitude decay at 1023 K of two profiles in 
orthogonal orientations on the 5$^\circ$  miscut Au(111) surface. The 
substantial difference in decay rates illustrates the predominant influence of 
step and step interaction energy, respectively.

 Strictly speaking, since 
there are unstable orientations near the (111) 
facet of Au, the theory by Bonzel and Mullins is not 
valid. This case will be addressed in section \ref{ch11}.

	Unfortunately, this kind of experiment always yields products of surface 
diffusion coefficient times surface energy term, and unravelling these two 
factors seems difficult. A combination of kinetic and equilibrium 
experiments offers a possibility to separate them but the total effort is 
substantial.

	Profile decay is exponential with time on vicinal surfaces for the small 
slope approximation~\cite{Bonzel2}, and the relaxation time $\tau$ is 
proportional to  
$\lambda^4$. This is the same result  (\ref{8.13})
as for a non-singular surface.

A numerical investigation {\it beyond} the small slope approximation
has also been done~\cite{Bonzel2} for the $\psi=\pi/2$ case. 
The calculated decay is 
linear with time. It also does no longer obey a $1/\sin\alpha$ dependence 
but rather 
saturates with decreasing miscut angle (Fig. \ref{neubild}). For $\psi=0$,
the corresponding profile 
shape is sinusoidal for small slopes and more trapezoidal when larger slopes 
are permitted. 
This blunting can be understood intuitively as follows.~\cite{Lancon} 
Near an isolated step,
$\delta \mu$ is proportional to the step curvature as seen from (\ref{GibbsT}).
Therefore, the steps try to decrease $\delta \mu$ by  increasing their 
radius of curvature, where it is most easy, namely at the top
and at the bottom, while in the intermediate region they are squeezed 
together, so that they cannot change their shape much. 
The increase of the radius of curvature results in a blunting of the profile.
For small amplitudes, the squeezing effect is absent and the profile 
remains sinusoidal. 
Another interpretation of the blunting is the following. 
The miscut has the effect to suppress the singularity of the free energy 
as a function of $z_x$ (\ref{gruber}), so that the Bonzel-Preuss~\cite{bonzel3}
calculation applies, which does predict a blunting.  
Simulations showed that increasing the step 
interaction energy 
relative to the step energy causes the steep portion between maximum and 
minimum of the profile to become less steep, as one expects. 

Keeping 
the step energy parameters constant, the influence of $\alpha$ on profile 
shape was also studied using the same algorithm. 
The smaller the 
miscut $\alpha$, the flatter the tops and bottoms of the profile. The 
behavior of 
the interesting $\psi=\pi/2$ case for larger slopes, where steps of 
opposite sign    
would be generated by the modulation, has not been dealt with so far. The 
question whether such profiles would facet or not remains unresolved. This 
case is, from an experimental point of view, an enormous challenge because 
the positioning of the profile parallel to the intrinsic steps would have to 
be very accurate.

Bonzel \& Mullins~\cite{Bonzel2} have also treated  the 
decay of a periodic profile of small amplitude  ($h/\lambda \ll \alpha$)
when there is a crossover from  migration-limited 
to  attachment-detachment surface-diffusion kinetics. The crossover 
occurs in the vicinity of a wavelength $\lambda^*$. The above relation
$\tau \propto \lambda^4$, characteristic of migration-limited kinetics,
holds only for $\lambda \gg \lambda^*$, while
$\tau \propto \lambda^2$, characteristic of attachment-detachment-limited kinetics,
holds only for $\lambda \ll \lambda^*$. This is a fairly remarkable result
which contrasts with the case of competition between surface diffusion kinetics
and evaporation-deposition kinetics on non-singular surfaces~\cite{Mullins}. 
Then, $\tau$ is proportional to
$\lambda^4$ for short wavelengths and to $\lambda^2$ for long wavelengths.


\section{Long range interactions}
\label{ch10}

To our knowledge, all computer simulations have been made without taking 
long range interactions into account, while  theoretical 
treatments~\cite{Rettori,Ozdemir,Spohn,Tang} 
use an entropic interaction in $1/\ell^2$ between {\it neighbouring} 
steps, which corresponds to a contact repulsion.

This is a questionable choice, because M\'etois~\cite{Alfonso} 
has experimentally 
shown that the elastic or electrostatic 
interaction is larger than the entropic interaction
even at fairly high temperature.

In the present section, we briefly discuss the expected effect of 
long range, repulsive interactions... and find quite disappointing results.

As seen in section \ref{sch3a}, the total elastic and electrostatic interaction
per unit length between two steps at distance $\ell$ is
$C/\ell^2$ for steps of identical sign, and $C'/\ell^2$
for steps of opposite sign, where $C$ and $C'$ are constants.

We shall first adress the case $C=C'>0$, neglecting step fluctuations. 
Then, the sign of the steps is irrelevant, and therefore a state 
with equally distant steps  (Fig. \ref{interact})
is an equilibrium state, since it is an 
equilibrium state when all steps have the same sign. For appropriate 
step signs, one can give to this equilibrium state any arbitrary height $h$
and wavelength $\lambda$.  
The existence of an equilibrium state of  arbitrary height 
and wavelength in the absence of step fluctuations suggests a very slow
decay when step fluctuations are taken into account.
The profile has a zigzag shape without facets. This
picture is very far from experiments.

Can we do better by assuming $C>C'>0$? It is easily seen 
from a perturbative treatment that there is still 
an equilibrium state of  arbitrary height 
and wavelength in the absence of step fluctuations, at least if $C'$
is not too different from $C$. The profile has 
again a zigzag shape without facets. The tops and bottoms are even sharper.
We do not get closer to experiments although the fluctuation-induced 
relaxation can be somewhat faster.

In the case $C'>C>0$, the perturbative treatment shows that there is also 
an equilibrium state of  arbitrary height, but now the profile has 
facets. However, an extremely slow 
relaxation is expected.

These results are quite disappointing. On one hand, steps should 
interact. On the other hand, taking interactions into account yields
results which disagree with experiments.
This paradox disappears if the surface is miscut.


\section{Unstable orientations}
\label{ch11}

In section \ref{ch2}, we mentioned that vicinal (001) and (111)
faces of Au are thermodynamically unstable. This property may be related
to an interaction between steps which is not uniformly repulsive
in contrast with the assumptions made, for instance in section \ref{ch4}.
The origin of this interaction may be one of those addressed in subsection
\ref{sch3b}.
However, although it is well known that  Au(111) 
	possesses a $22 \times \sqrt{3}$ reconstruction, the role played by the surface 
	reconstruction is not so clear as in the case of vicinal Au(100) 
	surfaces or vicinal Si(111) surfaces.
	`'Magic vicinals'' have been invoked, both theoretically~\cite{Bartolini} 
	and experimentally, in order to explain the remarkable stability of 
	particular vicinals. Such surfaces are characterised by a strong 
	relationship between the terrace width and the reconstruction mesh. 
	It has been observed for example~\cite{Samson,Sotto} on vicinals Au(100) and also~\cite{Watson,Yoon} on vicinal Pt(100).
	The role of surface reconstruction is clearly evidenced in 
	the pioneering work of Williams et al.~\cite{Williams} on Si(111), 
which is reported in subsection \ref{sch2}.

	As pointed out initially by Marchenko~\cite{Marchenko'} and more explicitly by 
	E. Williams and her colleagues~\cite{Williams}, elastic effects
are important in the ``phase separation'' phenomenon described in subsection
\ref{sch2}. Without taking elasticity into account, an unstable surface,
e.g. Au(11 9 9) would separate into domains whose size would increase 
indefinitely with time. 

	Elasticity limits the size of the domains,
	because each surface element of each ``phase'' exerts a stress on the solid, 
	the result of which is an elastic interaction proportional~\cite{Landau} to 
	$1/|\vec{r}-\vec{r}'|^3$ 
	between two surface elements $d^2r$ and $d^2r'$ at points $\vec{r}$
	and $\vec{r}'$. 
	The total energy of a stripe of size $\ell$
	is obtained by summing over the two coordinates of the points atoms 
	$\vec{r}$ and $\vec{r}'$. This operation multiplies by the
	sample width $L_y$ and introduces 3 integrals which transform the function
	$1/|\vec{r}-\vec{r}'|^3$ into a divergent function of $\ell$ proportional to
	$\ln \ell$. Similarly, the energy of a periodic array of stripes of two 
	different orientations is proportional to  $L_y \ln \ell$ per period,
	so that the elastic energy density is proportional to $\ln \ell/\ell$.
	The coefficient turns out to be positive 
	if the average orientation is unstable~\cite{Marchenko',Williams}. 
	Since the energy density resulting from domain boundaries is proportional 
	to $1/\ell$, too large domains are thermodynamically
	unstable with respect to domain wall formation. To summarize,  an
	unstable surface can reach an equilibrium state by forming a periodic
	structure of alternate ``phases''. The value of the period is given by a
	balance in energy between the  elastic energy 
	and the energy cost of boundaries. 
Indeed, a long range
order is present on the stepped surfaces of Au(111) described previously in
subsection \ref{sch2} (figures 
\ref{sylvie1}
and \ref{sylvie2}). The morphology of the faceted Au(11,9,9)
surface (figure \ref{sylvie1}) is clearly periodic. 
The value of the period is about
65 \AA, as it was determined both using diffraction technique (LEED) and
statistical analysis of STM images. In the case of Au(5,5,4), the
quasi-period is about 1800 \AA. The factor of more than 20 between the values
of both surfaces can be qualitatively  understood as follows.

The surfaces of interest are vicinals of (111) defined by Miller indices
($mmp$), with $m/p$ close to 1. They all correspond to [1,-1,0] steps,
but these steps have a different structure for $m/p<1$ and for
$m/p>1$ (compare fig.\ref{sylvie1}a and \ref{sylvie1}b). 
Their energy can therefore be quite different. Moreover, since the 
elastic energy is proportional to $\ln \ell$, the equilibrium 
domain width $\ell_{eq}$ is an exponential function of the step energy, 
and this can explain the very high ratio between the widths in figures 
	fig.\ref{sylvie2}a and \ref{sylvie2}b.  
	Details are given by Rousset et al.~\cite{Roussetal} 
It should be pointed out that, so far, no evolution of the faceted
morphologies has been found by increasing, either the temperature of
annealing (since 400$^{\circ}$C until 700$^{\circ}$C), 
or the time of annealing (since a few
minutes until 24 hours). This suggests  that the observed morphologies are
close to the equilibrium state. Perhaps not {\it quite} the equilibrium state. 
	Indeed, the evolution toward the
	thermodynamically stable state can be expected to be very slow.
	This slow evolution results from the attractive interaction between steps which 
	has been seen to be presumably responsible for the instability.
	This interaction favours the formation of step bunches observed in Fig. \ref{sylvie1}.
	The motion of these bunches is presumably more difficult than that of isolated steps,
	just as big particles diffuse less easily in any medium than slow particles.
	Moreover, if the bunches are close to the optimal size $\ell_{eq}$, 
	the evolution toward this size requires breaking the bunches, which is certainly
	very difficult. More precisely, the evolution toward equilibrium 
	should be rather slow until the period reaches a value of order $\ell_{eq}/2$, and
	much slower yet afterwards. The above discussion holds for $mmp$ surfaces 
	with $m/p<1$, whereas the case $m/p>1$ is more complicated.
	
	The attraction between steps is also expected to produce a very slow decay   
	of grooves on Au(111) and Au(001)~\cite{Duport}.
These grooves are completely different from
those on Ni(111) and Ni(001), whose vicinals are stable. In particular, there
is no objection to the formation of facets since unstable orientations
result from attractive interactions between  steps
of identical sign, while the instability of facets, advocated by certain 
theorists~\cite{Rettori,Ozdemir}, is related to the
repulsive interaction. Moreover, these facets are expected to 
decay very slowly since the upper and lower steps are 
tightly bound to their colleagues and can but hardly detach from them
and combine with the opposite step of opposite sign.

Groove healing has also been observed~\cite{Surnev,Surnev2} 
on Au(111) vicinal surfaces 
($\alpha =1.5^{\circ}, 5^{\circ}$), where the 
decay of perturbations of orientations $\psi=0$ and $\psi=\pi/2$ were 
observed  ($\psi$ is defined in section \ref{ch9}).These profiles could 
not be treated  under the small slope 
approximation. Initially the etched profiles have nearly square wave shapes, 
but after some time of annealing, sharp edges are eliminated. The decay of the 
$\psi=0$  profiles is much faster than  the one for
 $\psi=\pi/2$ profiles. The former shows a well-rounded shape after some time 
 of annealing, while the $\psi=\pi/2$ profile does not become sinusoidal even 
 after a very long time.

For $\psi=\pi/2$  a (111) facet is present, 
in adition to  an extra facet which is flat bounded by edge-like features 
indicative of missing orientations. The  profile in this case  becomes 
very asymmetrical. A possible explanation is that, when the profile is etched 
on a vicinal surface, the terrace widths are not the same.
When the terrace widths are smaller than a given minimum, if the 
average orientation of this part of the profile is close to a magic
orientation, it is energetically more favorable for the surface 
to present a flat region than to maintain the structure of steps/small terraces.


\section{Conclusion}
\label{ch12}

We have reviewed the state of the art of groove smoothing.  

When the average surface is  non-singular and far from any singular orientation,
everything is clear and there is nothing to add to Mullins' theory.

When the average surface is singular and  the neighbouring orientations 
are unstable, the situation is pretty clear too. Facets should appear at the 
top and at the bottom, they have sharp edges,\cite{Duport,Liu} 
and the decay is 
expected to be  generally very slow. An experimental check of the theory
is not yet available, and can obviously
be made only if the decay is not too slow.


When the average surface is singular and  the neighbouring orientations 
are stable, e.g. Ni(001), the situation is more complicated, except for a miscut
surface .

Let the controversial state of the art be summarized in the case of a singular
surface of an ideal, perfectly cut crystal, when the neighbouring orientations
are stable. The most common theoretical approach is a generalization of
Mullins' theory, using the non-equilibrium chemical potential $\mu$. 
However, the choice of the appropriate formula for $\mu$ is controversial.
A correct theoretical treatment should correctly take into account 
the peeling of the successive layers, and therefore the fluctuations 
of the top and bottom ledges. The attempt presented in section \ref{ch8}
tends to confirm the results obtained without taking fluctuations into account,
i.e. there is no facet on a well-cut profile. However, the treatment of 
fluctuations in section \ref{ch8} is far from being exact. The most 
recent experiments suggest that 
facet do appear in well-cut surfaces as well. This is in agreement with 
certain theories~\cite{Spohn,Hager}, although these theories 
do not take fluctuations into account in a satisfactory way either.

A point which requires further investigation is the `transient' attraction
investigated in section \ref{ch7}. It is clearly effective for
an (unphysical) one-dimensional surface. It appears
between two steps of identical sign  because an atom diffusing from the lower 
to the upper step has a shorter path and therefore a lower probability 
to come back. It does produce a blunting of a one-dimensional profile.
Whether it produces a similar effect on real surfaces is still unclear.

Most of existing theoretical treatments of the groove problem ignore long
range (elastic or electric) interactions between steps. The attempt which 
was done here does not yield satisfactory results. However, these
interactions are extremely important in the other problem reviewed here, 
namely phase separation of vicinal surfaces. A careful comparison of both
types of experiments in the same materials would probably lead to a better
understanding of interactions between steps.

\newpage


\vskip1cm

\begin{figure}
\caption{ Stepped areas found on a Au(111) surface misoriented by 50'. Note
that steps are straight on the right and lower part of the figure, whereas
kinks are present on the left and upper part. All steps are monoatomic,
2.35\AA high. The size of the STM image is 1020\AA $\times$ 1020\AA. 
Light white lines, running perpendicular to the steps, are the sign 
of the $22 \times \sqrt{3}$ reconstruction.}
\label{fig1}
\end{figure}

\begin{figure}
\caption{ Interference micrographs of periodic profiles on Ni(110) and 
Ni(100) single crystal surfaces, both annealed at 1173 K for several 
hours. The profiles exhibit a wavelength of 12 micrometers and 
amplitudes of 0.21 and 0.27 micrometer, respectively. Note the 
completely different shapes in the two cases. 
From Yamashita {\it et al.}} 
\label{fig2}                   
\end{figure}       

\begin{figure}
\caption{  Interference micrograph of a periodic profile on a Au(111) 
surface annealed at 1123 K. The wavelength and amplitude are 7.0 and 
0.125 micrometer, respectively.  }
\label{gold}
\end{figure}       

\begin{figure}
\caption{Decomposition of an unstable surface: Au(11,9,9).
(a) Hard sphere model of the ideal Au(11,9,9). Note that the steps consist
of \{100\} microfacets, displayed in grey.
(b) STM image of the Au(11,9,9) faceted surface. The size of the image is
356\AA $\times$ 314\AA. All  steps are monoatomic (2.35\AA~~high) and 
aligned along the zone axis $<0,1,-1>$.
 }
\label{sylvie1}
\end{figure}

\begin{figure}
\caption{Decomposition of an unstable surface: Au(5,5,4).
a) Hard sphere model of the ideal Au(5,5,4). Note that the steps consist of
\{111\} microfacets, displayed in grey.
(b) STM image of the Au(5,5,4) faceted surface. The size of the image is
1.07 $\mu$ m $\times$ 1.07 $\mu$ m.
(c) zoom in of the frontier between both phases, and located at the white
spot on fig. 5b. On the left part, terraces are 13\AA~~wide, whereas on the
right part they are about 35\AA~~ wide.}
\label{sylvie2}
\end{figure}

\begin{figure}
\caption{a) A vicinal surface. b) Free energy as a function of the slope}
\label{fig6}
\end{figure}

\begin{figure}
\caption{a) Decay of grooves considered as arrays of straight steps.
b) Effect of fluctuations}
\label{rettovi}
\end{figure}

\begin{figure}
\caption{                                    
a) The one-dimensional surface of a two-dimensional crystal. b)
The same as in the previous figure when the height is just two layers}
\label{paires}
\end{figure}

\begin{figure}
\caption{ Log-plot of profile amplitude versus annealing time for two 
profiles of 4.3 micrometer etched on a 5$^{\circ}$  miscut Au(111) crystal in 
the $\psi=\pi/2$ and $\psi=0$ orientations. 
The ratio of decay rates at 1023 K is 
25. From  Surnev {\it et al.} }
\label{xxx}
\end{figure}

\begin{figure}
\caption{ Numerically calculated decay rate versus amplitude of 
periodic profiles with modulation $\psi=0$ for three different miscut 
angles $\alpha$. Parameters: step free energy = 0.4, step interaction 
energy = 0.16, wavelength = 4 . From Bonzel \& Mullins }
\label{neubild}
\end{figure}
                                            
\begin{figure}
\caption{A stable configuration (in the absence of fluctuations) 
in the case of 
steps interacting through a long range interaction independent of
the sign of the steps.}
\label{interact}
\end{figure}

\end{document}